# *Chrysanthemum-like high-entropy diboride nanoflowers: A new class of high-entropy nanomaterials*

Da Liu, Honghua Liu, Shanshan Ning, Yanhui Chu[*]

School of Materials Science and Engineering, South China University of Technology,

Guangzhou 510641, China

---

[*] Corresponding author. Tel.:+86-20-82283990; fax:+86-20-82283990.
 *E-mail address:* chuyh@scut.edu.cn (Y.-H. Chu)

*Abstract*

High-entropy nanomaterials have been arousing considerable interest in recent years due to their huge composition space, unique microstructure, and adjustable properties. Previous studies focused mainly on high-entropy nanoparticles, while other high-entropy nanomaterials were rarely reported. Herein, we reported a new class of high-entropy nanomaterials, namely $(Ta_{0.2}Nb_{0.2}Ti_{0.2}W_{0.2}Mo_{0.2})B_2$ high-entropy diboride (HEB-1) nanoflowers, for the first time. The formation possibility of HEB-1 was first theoretically analyzed from two aspects of lattice size difference and chemical reaction thermodynamics. We then successfully synthesized HEB-1 nanoflowers by a facile molten salt synthesis method at 1473 K. The as-synthesized HEB-1 nanoflowers showed an interesting chrysanthemum-like morphology assembled from numerous well-aligned nanorods with the diameters of 20-30 nm and lengths of 100-200 nm. Meanwhile, these nanorods possessed a single-crystalline hexagonal structure of metal diborides and highly compositional uniformity from nanoscale to microscale. In addition, the formation of the as-synthesized HEB-1 nanoflowers could be well interpreted by a classical surface-controlled crystal growth theory. This work not only enriches the categories of high-entropy nanomaterials but also opens up a new research field on the high-entropy diboride nanomaterials.

*1. Introduction*

High-entropy materials, including high-entropy alloys and ceramics, that consist of four or more principal elements in near-equiatomic ratios or at least with each elements being between 5 and 35 at.% have attracted significant attentions since 2004 owing to their huge composition space, unique microstructure, and adjustable properties.[1,2] In the past decade, the research of high-entropy materials mainly focused on high-entropy alloys. Extensive experimental studies have demonstrated that high-entropy alloys exhibit many impressive properties, such as superior strength and ductility, good corrosion resistance, interesting creep characteristics, and so on.[3-5] Inspired by high-entropy alloy concept, the newly developed field of high-entropy ceramics has been gaining growing attention in recent years for potential applications in the structural and functional fields, especially extreme conditions of temperature, pressure and others. Up to date, numerous attempts have been made to exploring a variety of high-entropy ceramics, including metal oxides,[2,6,7] carbides,[8-11] and diborides.[12-15] with many superior physical and chemical performances, such as high hardness, low thermal conductivity, good thermodynamics stability and corrosion resistance, superior electrochemical and physicochemical properties, and so on. Compared with their bulk counterparts, high-entropy nanomaterials, including high-entropy nanoparticles, nanorods/wires/tubes, nanoplates, nanoflowers, etc., have much more unique physical and chemical characteristics due to their five core effects including small size effect, surface effect, quantum size effect, quantum tunneling effect, and dielectric confinement effect, which makes them exhibit the promising applications in nanophotonics, nanoelectronics, nanocomposites, chemical and biological sensing, and so on.[16,17] Therefore, the recent research work has begun to focus on the synthesis of high-entropy nanomaterials. Nevertheless, only a couple of high-entropy alloy and ceramic nanoparticles have been successfully explored until now.[18,19] Extending the state of the art for new classes of high-entropy nanomaterials is still a great challenge for the scientific community.

High-entropy diborides, as a number of high-entropy ceramics, are very attractive owing to the unusual combination of the attractive physico-chemical properties for their diboride components, such as melting temperature exceeding 3000 K, high hardness, chemical inertness, good electrical and thermal conductivity, low neutron absorption, etc..[12-15] Among the existing nanostructures, the nanoflowers possess the most promising applications in nanophotonics, nanoelectronics, nanocomposites, and chemical and biological sensing.[20] In this work, we open up new opportunities for research of a new class of high-entropy nanomaterials, namely $(Ta_{0.2}Nb_{0.2}Ti_{0.2}W_{0.2}Mo_{0.2})B_2$ high-entropy diboride (HEB-1) nanoflowers. We first analyzed the formation possibility of HEB-1 nanoflowers from two aspects of lattice size difference and chemical reaction thermodynamics and then successfully synthesized the chrysanthemum-like HEB-1 nanoflowers by a facile molten salt synthesis method at 1473 K. The phase compositions, morphology and compositional uniformity of the as-synthesized HEB-1 nanoflowers were investigated, as well as their growth mechanisms. To the best of our knowledge, it is the first time that the high-entropy nanoflowers are synthesized and investigated. This work not only enriches the categories of high-entropy nanomaterials but also will explore a new research field in the high-entropy diboride nanomaterials.

*2. Experimental procedure*

The commercially available $Nb_2O_5$ and $Ta_2O_5$ powders (99.9% purity, particle size: 1-3 μm, Shanghai ChaoWei Nanotechnology Co. Ltd., Shanghai, China), $TiO_2$, $WO_3$ and $MoO_3$ powders (99.9% purity, particle size: 100-300 nm, Shanghai ChaoWei Nanotechnology Co. Ltd., Shanghai, China), and amorphous B powders (99.9% purity, average particle size < 3 μm, Shanghai ChaoWei Nanotechnology Co. Ltd., Shanghai, China) were used as the starting materials. NaCl/KCl were used as molten salt medium. Details of the synthesis of HEB-1 nanoflowers were described as follows: The aforementioned starting materials and NaCl/KCl salts were first milled by hand for 30 min in an agate mortal using an agate pastel based on the following ratios: 1:1:2:2:2:35 for $Ta_2O_5/Nb_2O_5/TiO_2/WO_3/MoO_3/B$ (molar ratio), 1:1 for NaCl/KCl (molar ratio), and 10:1 for (NaCl, KCl)/($Ta_2O_5$, $Nb_2O_5$, $TiO_2$, $WO_3$, $MoO_3$, B) (weight ratio). The mixture of above powders was put into an alumina crucible and then placed into a horizontal alumina tube furnace. Afterwards, the furnace was heated at a rate of 10 K/min to desired temperature and then maintained for 30 min. The whole heating and cooling process was carried out in flowing Argon gas. After cooling naturally to room temperature, the as-synthesized products were taken out, immersed in the deionized water at 363 K and the absolute ethanol at 273 K, and filtered onto Nuclepore filters to remove the residual NaCl/KCl salts and $B_2O_3$ products. Finally, they were repeatedly washed, filtered and dried at 333 K. In addition, to investigate the growth mechanism of HEB-1 nanoflowers, the additional experiments were conducted at 1473 K for different holding time (10 min, 20 min and 30 min).

The phase compositions of the as-synthesized products were analyzed by X-ray diffraction (XRD, X'pert PRO; PANalytical, Almelo, Netherlands). The equivalent counting time for a conventional point detector would be 30 s per point at 0.01° 2θ increments. The microstructure and compositional uniformity of the as-synthesized products were characterized by scanning electron microscopy (SEM, Supra-55; Zeiss, Oberkochen, Germany) equipped with energy dispersive spectroscopy (EDS) and transmission electron microscopy (TEM, Tecnai F30G2; FEI, Eindhoven, Netherlands) equipped with EDS.

## 3. Results and discussion

To analyze the synthesis possibility of HEB-1 products by molten salt synthesis method, a critical empirical parameter, lattice size difference ($\delta$), that is usually calculated to predict the formation ability of solid solutions is first analyzed and it can be expressed in high-entropy diborides by the following equation:[21]

$$\delta = \sqrt{\sum_{i=1}^{n} \frac{n_i}{2}\left[(1-\frac{a_i}{\bar{a}})^2 + (1-\frac{c_i}{\bar{c}})^2\right]} \qquad (1)$$

where $n$ is the metal diboride component species in high-entropy diborides, $n_i$ is the molar fraction of the $i$th MeB$_2$ component of high-entropy diborides. $a_i$ and $c_i$ are the corresponding lattice parameters of the individual metal diborides, and $\bar{a} = \sum_{i=1}^{n} n_i a_i$ and $\bar{c} = \sum_{i=1}^{n} n_i c_i$ are the average lattice parameters, respectively. In general, a smaller $\delta$ indicates a smaller lattice size difference and finally benefits the formation of solid solutions. According to the reported lattice parameters of individual metal diborides,[13,22] the $\delta$ value of HEB-1 can be calculated to be 2.763%, less than that of the previously reported $(Hf_{0.2}Zr_{0.2}Ta_{0.2}Nb_{0.2}Ti_{0.2})B_2$ high-entropy diborides (3.109%),[13] which suggests that the HEB-1 products can be synthesized by molten salt synthesis method from the aspect of lattice size difference. In addition, the thermodynamics of the possible chemical reactions for the starting materials is further analyzed. In our work, the starting materials mainly consist of $TiO_2$, $Nb_2O_5$, $Ta_2O_5$, $WO_3$, $MoO_3$ and B powders, while NaCl and KCl salts are only utilized as the molten salt medium. Therefore, on the basis of the reported thermodynamic data of the individual metal diborides,[23] the possible reactions as well as the correlations between the standard Gibbs free energy ($\Delta G^{\theta}_{R,T}$/kJ·mol$^{-1}$) of these starting materials and the temperature (T/K) can be described as follows:

$$TiO_2 + \tfrac{10}{3}B \rightarrow TiB_2 + \tfrac{2}{3}B_2O_3(l) \qquad (2)$$

$$\Delta G^{\theta}_{R,T} = -223930 + 4T$$

$$\tfrac{1}{2}Nb_2O_5 + \tfrac{11}{3}B \rightarrow NbB_2 + \tfrac{5}{6}B_2O_3(l) \qquad (3)$$

$$\Delta G^{\theta}_{R,T} = -405640 + 5T$$

$$\tfrac{1}{2}Ta_2O_5 + \tfrac{11}{3}B \rightarrow TaB_2 + \tfrac{5}{6}B_2O_3(l) \qquad (4)$$

$$\Delta G^{\theta}_{R,T} = -240372 + 2T$$

$$WO_3 + 4B \rightarrow WB_2 + B_2O_3(l) \qquad (5)$$

$$\Delta G^{\theta}_{R,T} = -672560 - 22T$$

$$MoO_3 + 4B \rightarrow MoB_2 + B_2O_3(l) \qquad (6)$$

$$\Delta G^{\theta}_{R,T} = -603740 - 17T$$

$$\tfrac{1}{5}TaB_2 + \tfrac{1}{5}NbB_2 + \tfrac{1}{5}TiB_2 + \tfrac{1}{5}WB_2 + \tfrac{1}{5}MoB_2 \rightarrow (Ta_{0.2}Nb_{0.2}Ti_{0.2}W_{0.2}Mo_{0.2})B_2 \qquad (7)$$

$$\Delta G^{\theta}_{R,T} = \Delta G^{M}_{R,T}$$

$$\tfrac{1}{5}TiO_2 + \tfrac{1}{10}Ta_2O_5 + \tfrac{1}{10}Nb_2O_5 + \tfrac{1}{5}WO_3 + \tfrac{1}{5}MoO_3 + \tfrac{41}{15}B \rightarrow$$
$$(Ta_{0.2}Nb_{0.2}Ti_{0.2}W_{0.2}Mo_{0.2})B_2 + \tfrac{13}{15}B_2O_3(l) \qquad (8)$$
$$\Delta G^{\theta}_{R,T} = -419250 - 5T + \Delta G^{M}_{R,T}$$

where $\Delta G^{M}_{R,T}$ is the mixing Gibbs free energy of HEB-1. In this case, the HEB-1 is assumed to be an ideal Raoultian solution and thereby $\Delta G^{M}_{R,T}$ can be calculated by the following equation:

$$\Delta G^{M}_{R,T} = -T\Delta S_{mix} \qquad (9)$$

where $\Delta S_{mix}$ is the mixing entropy of HEB-1, which can be defined as:[13,14]

$$\Delta S_{mix} = -\tfrac{R}{3}\sum_{i=1}^{N} x_i \ln(x_i) \qquad (10)$$

where $R$ is the ideal gas constant, $N$ is the metal element species, and $x_i$ are the molar fractions of the $i$th metal element in the sublattices, respectively. According to the equation (10), the $\Delta S_{mix}$ of HEB-1 is calculated to be about 0.54R. As a result, the $\Delta G^{\theta}_{R,T}$ of the reactions (7) and (8) can be mathematically expressed as:

$$\Delta G^{\theta}_{R,T} = -5T \qquad (11)$$

$$\Delta G^{\theta}_{R,T} = -419250 - 10T \qquad (12)$$

In such a case, the thermodynamics analysis of the possible chemical reactions for the starting materials is depicted in Fig. 1(a). It can be observed that the standard Gibbs free energy of all the reactions ((2)-(8)) is negative ($\Delta G^{\theta}_{R,T} < 0$), and therefore they can all proceed spontaneously. Nevertheless, it is worth noticing that the reaction (7) is very difficult one to occur when it competes with other reactions due to its Gibbs free energy close to zero. But the reaction (8) is still prone to occur to generate HEB-1 products in the system owing to its very negative Gibbs free energy. That is to say, the synthesis of HEB-1 products via the reaction (8) using molten salt synthesis method is possible from the thermodynamic aspect. On the basis of the aforementioned theoretical analysis, we preformed the molten salt synthesis of HEB-1 products at different temperatures, and XRD characterization was conducted to determine the phase compositions of the as-synthesized products at different temperatures. In order to observe the weak diffraction peaks more clearly, X-ray diffraction data was plotted on a logarithmic scale, as displayed in Fig. 1(b). Clearly, the as-synthesized products at 1273 K are composed of a dominant (Ta, Nb, Ti, W, Mo)$B_2$ phase and a minor secondary phase. According to the JCPDS cards of NbB phase (JCPDS card No. 81-0911) and TaB phase (JCPDS card No. 81-0912). the minor secondary phase can be indexed to be (Ta,Nb)B phase that possesses different crystal structure with major high-entropy diboride phase. With the increase of the synthesis temperatures, the diffraction of minor (Ta, Nb)B phase gradually decreased and finally disappeared at 1423 K to form a single (Ta$_{0.2}$Nb$_{0.2}$Ti$_{0.2}$W$_{0.2}$Mo$_{0.2}$)$B_2$ phase with a hexagonal crystal structure of metal diborides. Therefore, the pure HEB-1 products can be successfully synthesized by molten salt synthesis method at 1423 K.

Fig. 2(a) presents low-magnification SEM image of the as-synthesized HEB-1 products. It can be observed that the as-synthesized HEB-1 products exhibit an interesting chrysanthemum-like morphology assembled from numerous nanorods. This flower morphology is similar to that of the reported as-synthesized TaB$_2$ and (Ta$_{1/3}$Nb$_{1/3}$Ti$_{1/3}$)B$_2$ solid solution via molten salt synthesis method,[21,24] but is different from the reported as-synthesized particle-like (Hf$_x$Zr$_{1-x}$)B$_2$ solid solutions via molten salt synthesis method.[25] This indicates that the multi-component solid solutions cannot contribute to the formation of the nanoflowers. High-magnification SEM image (Fig. 2(b)) shows that each individual chrysanthemum-like structure is composed of several dozen well-aligned nanorods. Those nanorods possess the low aspect ratio with the diameters of 20-30 nm and lengths of 100-200 nm. Fig. 2(c) displays EDS compositional maps of the as-synthesized HEB-1 nanoflowers (labeled by a dotted green square in Fig. 2(a)) at micrometer scale. Obviously, the distribution of five metal elements is highly homogeneous at micrometer scale and no segregation or aggregation is found throughout the scanned area, which implies that the as-synthesized HEB-1 nanoflowers possess the highly compositional uniformity at micrometer scale.

Fig. 3(a) is a typical TEM image of an individual HEB-1 nanoflower, which clearly displays that the as-synthesized nanoflowers are composed of several dozen well-aligned nanorods with the diameters of 20-30 nm and lengths of 100-200 nm on a nanocluster root. The high-resolution transmission electron microscopy (HRTEM) images (Fig. 3(b) and Fig. 3(c)) of two selected nanorods in Fig. 3(a) show that these nanorods possess a periodic lattice structure with two sets of fringes with the d-space of 0.205 nm and 0.113 nm in Fig. 3(b) or 0.326 nm and 0.268 in Fig. 3(c), corresponding to the $\{101\}$ and $\{\bar{2}12\}$ planes or $\{001\}$ and $\{\bar{1}10\}$ planes of $(Ta_{0.2}Nb_{0.2}Ti_{0.2}W_{0.2}Mo_{0.2})B_2$, respectively. The corresponding Fast Fourier transform (FFT) patterns (inserted in Fig. 3(b) and Fig. 3(c)) display that those nanorods are the single-crystal hexagonal structure of $(Ta_{0.2}Nb_{0.2}Ti_{0.2}W_{0.2}Mo_{0.2})B_2$. The HRTEM images, together with the corresponding FFT patterns, strongly demonstrate that these nanorods are single-crystal hexagonal structure of $(Ta_{0.2}Nb_{0.2}Ti_{0.2}W_{0.2}Mo_{0.2})B_2$. Meanwhile, the growth direction of the nanorods presented in Fig. 3(b) and Fig. 3(c) can be determined as <112> and <001>, respectively, which suggests that these nanorods grow up along the different directions. In addition, an amorphous layer of 2-3 nm can be observed on the surface of nanorods, as shown in Fig. 3(b) and Fig. 3(c). Fig. 3(d) shows the scanning transmission electron microscopy (STEM) image and the corresponding EDS compositional maps of two individual nanorods at nanometer scale. It is evident that the distribution of five metal elements in the nanorods is highly uniform at nanoscale without segregation or aggregation phenomenon and simultaneously their atomic percentages are the equal atomic quantity (Table 1), which further confirms that the as-synthesized nanoflowers are composed of $(Ta_{0.2}Nb_{0.2}Ti_{0.2}W_{0.2}Mo_{0.2})B_2$ phase. In addition, a small amount of O element can be detected by EDS, as listed in Table 1, which can account for the presence of the amorphous $B_2O_3$ layer on the nanorod surface.

In order to investigate the growth mechanism of the as-synthesized HEB-1 nanoflowers, the additional experiments were conducted at 1473 K for different holding time (10 min, 20 min and 30 min). XRD patterns and SEM images of the as-synthesized products were displayed in Fig. 4. As presented in Fig. 4(a), all the as-synthesized products possessed the same phase compositions and they were all composed of $(Ta_{0.2}Nb_{0.2}Ti_{0.2}W_{0.2}Mo_{0.2})B_2$ phase without any other phase. However, the as-synthesized products exhibited the different morphology at different holding time. When the holding time was 10 min, the as-synthesized products showed many clusters with diameter of several hundred nanometers involving numerous spherical nanoparticles, as shown in Fig.4(b). With the holding time increasing to 20 min, numerous nanorods with the diameters of 20-30 nm and lengths of 50-150 nm grew on the formed cluster surface, as displayed in Fig.4(c). As the holding time increased to 30 min, the as-synthesized products possessed the chrysanthemum-like morphology assembled from numerous well-aligned nanorods, as presented in Fig. 4(d). These results can well demonstrate the growth process of the as-synthesized HEB-1 nanoflowers. In other words, the formation of the as-synthesized HEB-1 nanoflowers underwent a transformation from nanoclusters to nanoflowers.

On the basis of the above-mentioned XRD, SEM and TEM analysis, the possible growth process of the as-synthesized HEB-1 nanoflowers was proposed and the corresponding schematic diagram was depicted in Fig. 5. In our work, the system consisted of five metal oxides, B powders and NaCl/KCl salts, as presented in Fig. 5(a). At elevated temperature, the NaCl/KCl salts would first melt into the liquid phase as the molten salt medium and then five metal oxides ($MeO_x$) dissociated to mobile metal cations ($Me^{x+}$) and $O^{2-}$ in the molten salt medium by the following equation:[26]

$$MeO_x = Me^{x+} + O^{2-} \qquad (13)$$

Meanwhile, B powders would also dissolve into the molten salt medium to form ions of $B^{3+}$ and electrons as the following equation,[26] as presented in Fig. 4(b).

$$B = B^{3+} + 3e^- \qquad (14)$$

Nevertheless, the solubility of B is very low in NaCl/KCl molten salt medium, which is less than that of metal oxides in NaCl/KCl molten salt medium.[26] In this situation, the metal cations and $O^{2-}$ would rapidly diffuse to the surface of B cations and reacted with each other to generate HEB-1 molecules and $B_2O_3$ by-products via the reaction (14), as illustrated in Fig. 5(c).

$$Me^{x+} + 3O^{2-} + 4B^{3+} + e^- = MeB_2 + B_2O_3 \qquad (15)$$

where $MeB_2$ presents HEB-1. This further confirmed the thermodynamics analysis of chemical reactions for the as-synthesized HEB-1 products. With the prolongation of the reaction time, the concentration of the generated HEB-1 molecules in molten salt medium would reach the supersaturation condition. Under the circumstances, the nucleation of HEB-1 would occur to generate numerous HEB-1 nucleus. Afterwards those nucleus would grow up to be the nanoflowers. In the following growth process, the formation of HEB-1 nanoflowers could be well interpreted by a classical interface-controlled growth instead of diffusion-controlled growth in molten salt medium based on Ostwald Ripening theory.[27,28] In general, the interface-controlled growth of crystal in solutions could be classified into two broad categories: rough interface-controlled growth and smooth surface-controlled growth.[29] The rough interface-controlled growth mechanism is generally utilized to interpret the growth of regular or spherical crystals, in which molecules can attach to the crystal surface at essentially any site to allow the interface to advance uniformly. While the smooth surface-controlled growth, which involves that molecules can only attach at the step to allow the growth to occurs at a particular place, is usually used to explain the growth of a variety of oriented and hierarchical crystals. Whether rough or smooth interface-controlled growth models can occur mainly depends on the intrinsic structure of materials and growth conditions. The Jackson $\alpha$-factor is usually proposed to predict whether rough or smooth interface-controlled growth models can occur from the aspect of material's structure and it can be expressed as the following equation:[30]

$$\alpha = \frac{\Delta H}{RT}\frac{\eta_1}{Z} \qquad (16)$$

where $\Delta H$ is enthalpy changes, $\eta_1/Z$ is the fraction of the nearest neighbor sites which are in a single layer of molecules at the surface, $T$ is the temperature, and $R$ is the gas constant. The growth model of crystal in solutions cannot be governed by the smooth interface-controlled growth mechanism unless the $\alpha$ value of the materials is greater than 2 and simultaneously the supersaturation of the system is under a moderate or low condition.[31] For metal diborides, the $\alpha$ values of main planes, such as {001}, {100} and {110}, are greater than 4,[32] from which it can be inferred that the $\alpha$ values of the planes for high-entropy metal diborides are also greater than 4. As a result, these two surface-controlled growth models of crystal in solutions mainly depends on the growth conditions of the system. At the initial stage, the concentration of HEB-1 molecules in molten salt medium is expected to be constant under a high supersaturation condition. Such condition will favor the rough interface-controlled crystal growth. In other words, the formed HEB-1 nucleus in molten salt medium will undergo the rough interface-controlled growth to form the individual nanoparticles with spherical morphology,[33,34] as displayed in Fig. 5(c). Those individual nanoparticles are not stable in molten salt medium due to their high specific surface energy and they will be aggregated by strong van der Waals interaction among them to form the nanoparticle clusters,[35] as shown in Fig 5.(d). With the further prolongation of the reaction time, the concentration of HEB-1 molecules in molten salt medium will decrease to a moderate or low supersaturation condition. In this situation, the following growth of the nanoparticle clusters will be governed by the smooth interface-controlled growth. It is well known that the smooth interface-controlled growth includes two-dimensional (2D) nucleation growth and screw dislocation growth.[36] According to SEM and TEM observations, the following growth of the nanoparticle clusters should be governed by the smooth interface-controlled 2D nucleation growth. According to the 2D nucleation growth theory, the growth rate ($v_c$) at the nucleation plane can expressed as:[37]

$$v_c = av\sigma \cdot \exp\left(\frac{E_{des}-E_{sd}}{2kT}\right)\exp\left(-\frac{\Delta U}{kT}\right) \qquad (17)$$

where $a$ is the distance between adsorption sites, $v$ is the vibration frequency, $\sigma$ is the supersaturation, $k$ is the Boltzmann constant, $T$ is temperature, $\Delta U$ is the active energy, $E_{des}$ is the activation energy for the desorption of an admolecule from the crystal surface and $E_{sd}$ is the activation energy for surface diffusion. For high-entropy metal diborides, the {100} plane possessed the highest surface desorption energy ($E_{des}$) among all the crystal facets.[38] Consequently, the growth velocity of the <100> direction should be the highest among all directions in thermodynamics. However, in a realistic crystal growth process, some environmental factors of the system, such as impurity pinning site, adsorbed surfactants, and transport kinetics, are complicated and varied, which can also result in the highest growth velocity in other directions in addition to <100> direction.[39] This is why the growth directions of the nanorods in HEB-1 nanoflowers are difference, as displayed in Fig. 3(b) and Fig. 3(c). In brief, under a moderate or low supersaturation condition, HEB-1 molecules in molten salt medium will nucleate on the surface active positions of the formed nanoclusters to decrease the system energy and simultaneously they will grow along the direction of the lowest energy based on the smooth interface-controlled 2D nucleation growth. Finally, the complicated and varied environmental factors resulted in the formation of HEB-1 nanoflowers with nanocluster roots based on the smooth interface-controlled 2D nucleation growth, as displayed in Fig. 5(f).

## *4. Conclusion*

In summary, a new class of high-entropy nanomaterials, namely HEB-1 nanoflowers, was reported in our work for the first time. We first theoretically analyzed the formation possibility of HEB-1 from two aspects of lattice size difference and chemical reaction thermodynamics. The HEB-1 nanoflowers then were successfully synthesized by a facile molten salt synthesis method. The as-synthesized HEB-1 nanoflowers exhibited the chrysanthemum-like morphology assembled from numerous well-aligned nanorods with the diameters of 20-30 nm and lengths of 100-200 nm. These nanorods had a single-crystalline hexagonal structure of metal diborides and simultaneously exhibited highly compositional uniformity from nanoscale to microscale. On the basis of SEM and TEM analyses, a classical surface-controlled crystal growth theory was proposed to interpret the formation of the as-synthesized HEB-1 nanoflowers.

*Acknowledgements*

The authors acknowledge financial support from the National Key Research and Development Program of China (No. 2017YFB0703200), National Natural Science Foundation of China (No. 51802100 and 51972116), and Young Elite Scientists Sponsorship Program by CAST (No. 2017QNRC001).

*Author Contributions*

Y. Chu conceived and designed this work. D. Liu, H Liu, and S Ning performed the experiments. Y. Chu and H Liu analyzed the data. All authors commented on the manuscript.

*Notes*

The authors declare no competing financial interest.

17. Chu Y, Jing S, Yu X, Zhao Y. High-temperature Plateau-Rayleigh growth of beaded SiC/SiO$_2$ nanochain heterostructures. Cryst Growth Des. 2018; 18(5): 2941-2947.

18. Yao Y, Huang Z, Xie P, Lacey SD, Jacob RJ, Xie H, Yu D, Hu L. Carbothermal shock synthesis of high-entropy-alloy nanoparticles. Science. 2018; 359(6383): 1489-1494.

19. Ning S, Wen T, Ye B, Chu Y. Low-temperature molten salt synthesis of high-entropy carbide nanopowders. J Am Ceram Soc. 2019; https://doi.org/10.1111/jace.16896.

20. Kharisov BI, A Review for Synthesis of Nanoflowers, Recent Pat Nanotech 2018; 2(3):190-200.

21. Wen T, Ning S, Liu D, Ye B, Liu H, Chu Y. Synthesis and characterization of the ternary metal diboride solid-solution nanopowders. J Am Ceram Soc. 2019; 102(8): 4956-4962.

22. Li P, Ma L, Peng M, Shu B, Duan Y. Elastic anisotropies and thermal conductivities of WB$_2$ diborides in different crystal structures: a first-principles calculation. J Alloy Compd, 2018; 747: 905-915.

23. Fahrenholtz WG, Wuchina EJ, Lee WE, Zhou Y. Ultra-high temperature ceramics: materials for extreme environment applications. John Wiley & Sons; 2014.

24. Wei T, Liu Z, Ren D, Deng X, Deng Q, Huang Q, Ran S. Low temperature synthesis of TaB$_2$ nanorods by molten-salt assisted borothermal reduction. J Am Ceram Soc. 2018; 101(1): 45-49.

*Table 1.* The elemental atomic percentage of the as-synthesized HEB-1 nanoflowers by STEM-EDS analyses.

| Elements | Ti | W | Ta | Nb | Mo | B | O |
|---|---|---|---|---|---|---|---|
| At. % | 6.13 | 6.02 | 5.77 | 5.63 | 5.50 | 69.94 | 1.01 |

*Figure captions*

*Fig. 1.* (a) Thermodynamics analysis of the possible chemical reactions for as-synthesized HEB-1 products; (b) XRD patterns of the as-synthesized products at different temperatures.

*Fig. 2.* SEM characterizations of the as-synthesized HEB-1 products: (a) Low-magnification SEM image; (b) high-magnification SEM image; (c) EDS elemental maps.

*Fig. 3.* TEM analyses of an individual HEB-1 nanoflower: (a) TEM image; (b) HRTEM image of the selected nanorod labeled by doted green square in (a); (c) HRTEM image of the selected nanorod labeled by doted blue square in (a); (d) STEM image and the corresponding EDS compositional maps.

*Fig. 4.* XRD and SEM characterizations of the as-synthesized products at 1423 K for different holding time: (a) XRD patterns; (b) SEM image of 10 min; (b) SEM image of 20 min; (d) SEM image of 30 min.

*Fig. 5.* Schematic diagram of the possible growth process for HEB-1 nanoflowers.

*Figures*

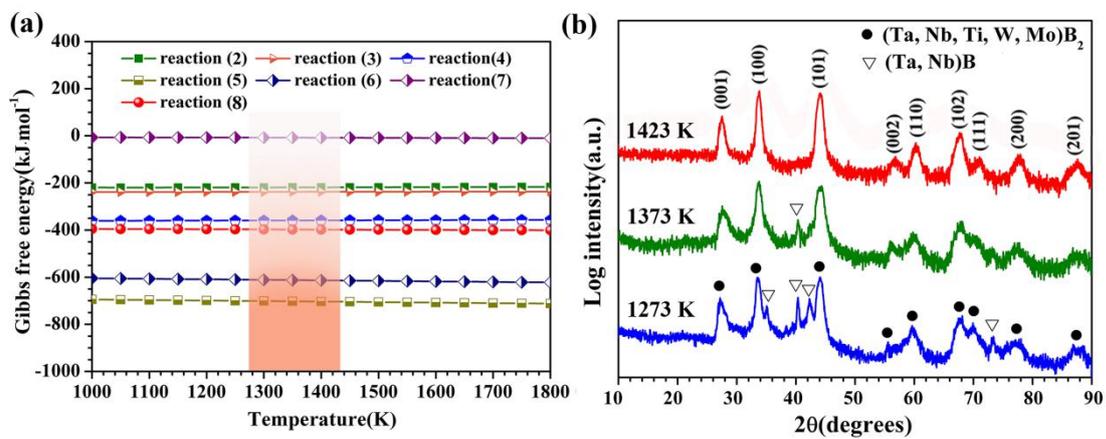

*Fig. 1.* (a) Thermodynamics analysis of the possible chemical reactions for as-synthesized HEB-1 products; (b) XRD patterns of the as-synthesized products at different temperatures.

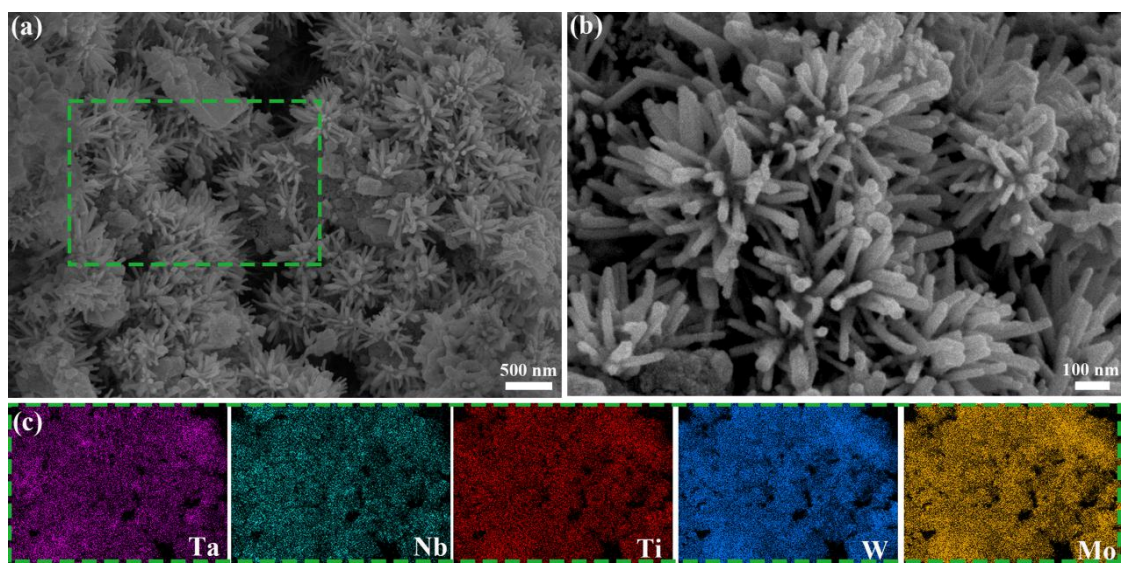

***Fig. 2.*** SEM characterizations of the as-synthesized HEB-1 products: (a) Low-magnification SEM image; (b) high-magnification SEM image; (c) EDS elemental maps.

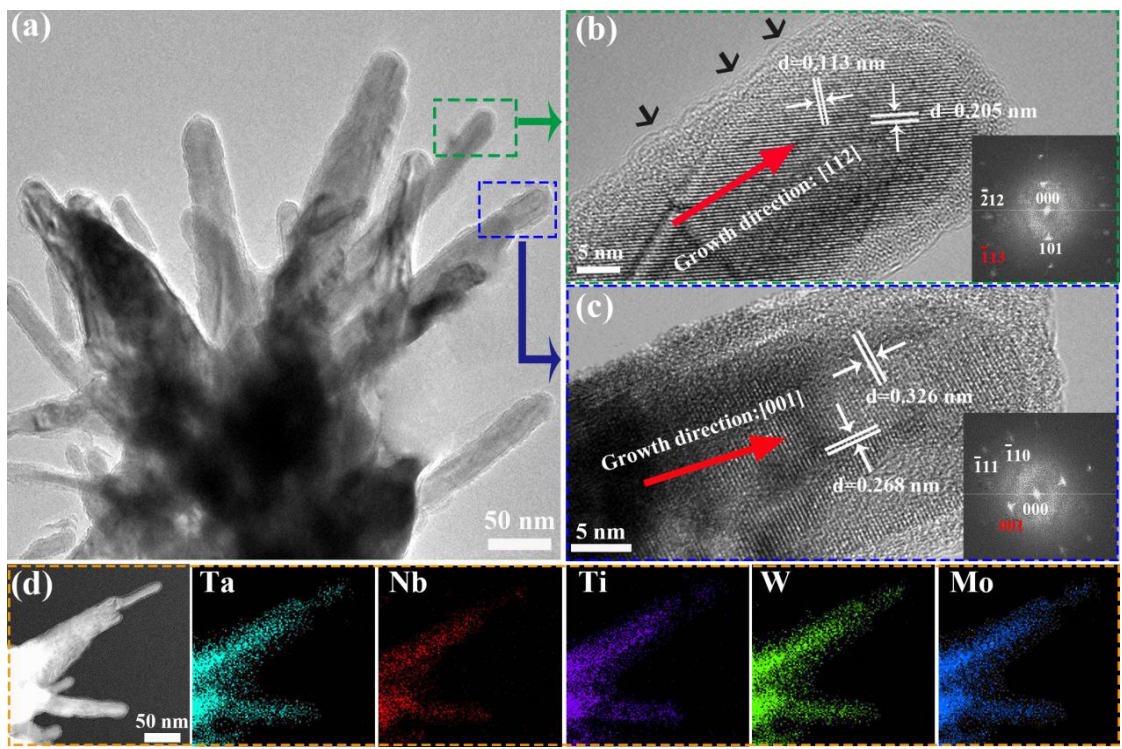

*Fig. 3.* TEM analyses of an individual HEB-1 nanoflower: (a) TEM image; (b) HRTEM image of the selected nanorod labeled by doted green square in (a); (c) HRTEM image of the selected nanorod labeled by doted blue square in (a); (d) STEM image and the corresponding EDS compositional maps.

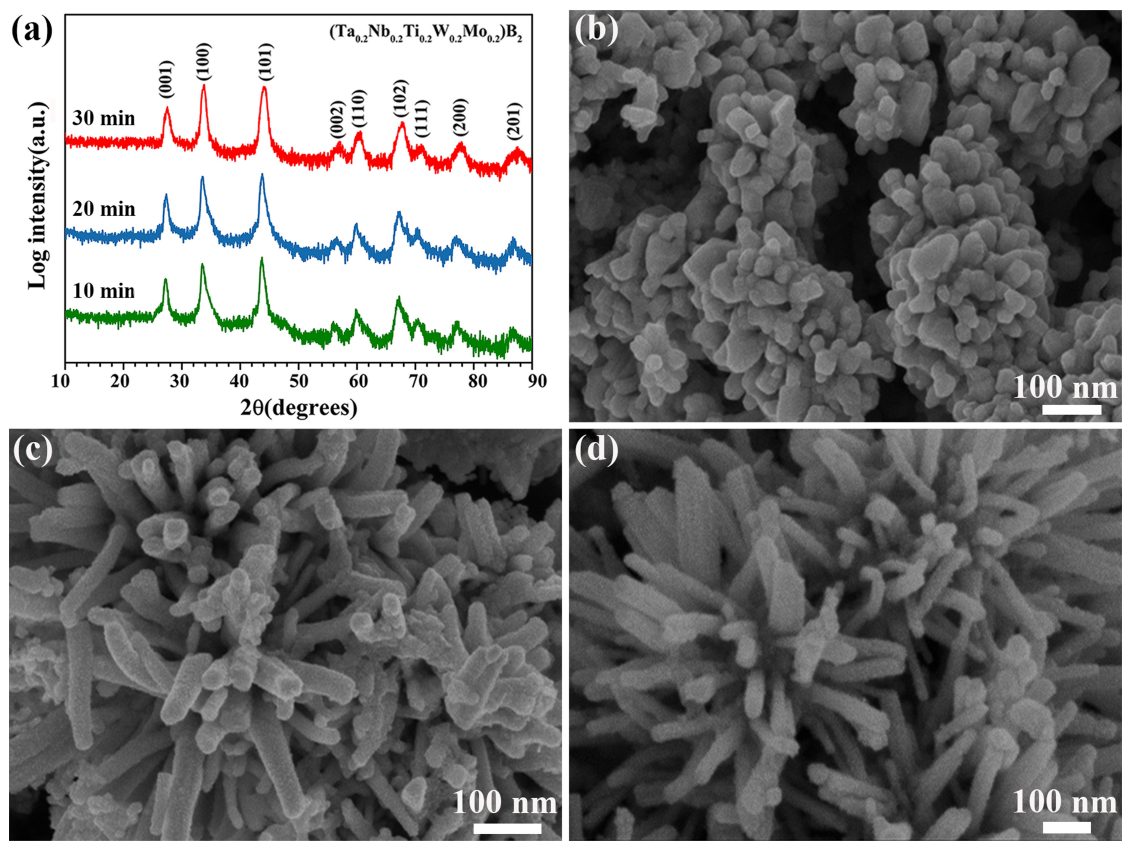

*Fig. 4.* XRD and SEM characterizations of the as-synthesized products at 1423 K for different holding time: (a) XRD patterns; (b) SEM image of 10 min; (b) SEM image of 20 min; (d) SEM image of 30 min.

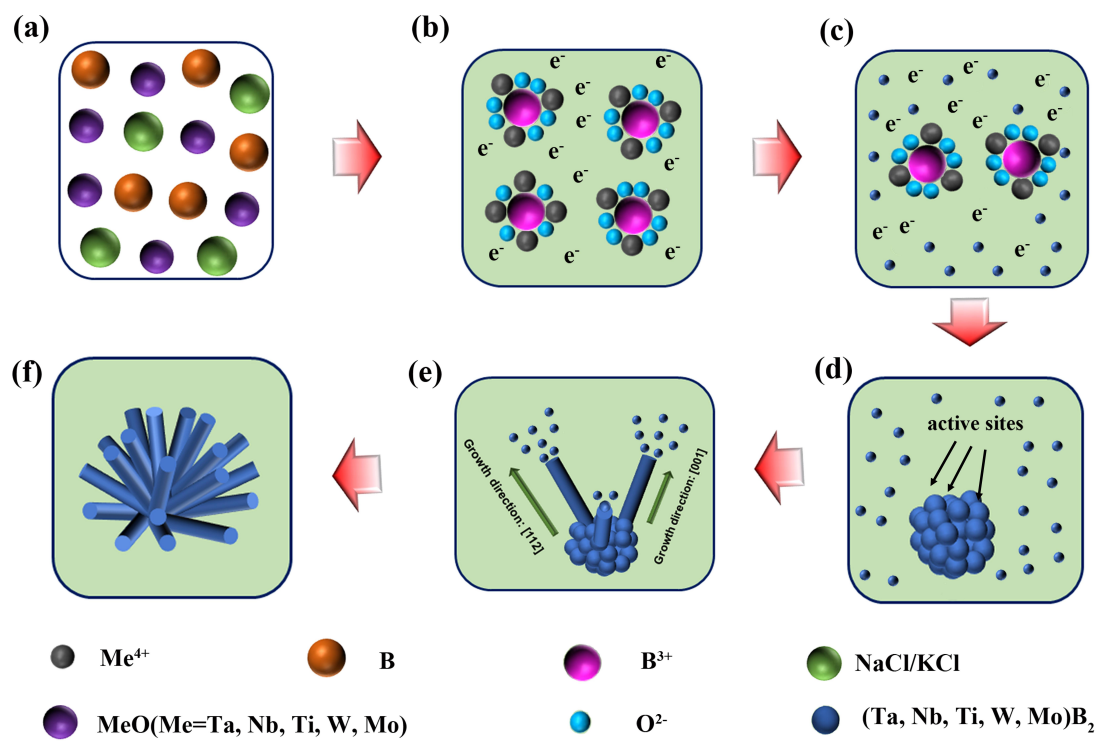

*Fig. 5.* Schematic diagram of the possible growth process for HEB-1 nanoflowers.